\documentclass[nofootinbib,aps,superscriptaddress]{revtex4}
\usepackage{amssymb,amsmath,graphicx,color,microtype}
\usepackage{appendix}
\usepackage{hyperref}
\hypersetup{
	colorlinks=true,
	linkcolor=blue,
	filecolor=blue,      
	urlcolor=blue, 
	bookmarks=true,
	citecolor=blue
}

\def\l{\left}
\def\r{\right}
\def\be{\begin{equation}}
\def\ee{\end{equation}}
\def\ba{\begin{eqnarray}}
\def\ea{\end{eqnarray}}
\def\bl#1\el{\begin{align}#1\end{align}}
\def\nn{\nonumber}

\begin{document}
	
\title{Cosmological Standard Timers from Unstable Primordial Relics}

\author{Yi-Fu Cai}
\email{yifucai@ustc.edu.cn}
\affiliation{Department of Astronomy, School of Physical Sciences, University of Science and Technology of China, Hefei, Anhui 230026, China}
\affiliation{CAS Key Laboratory for Research in Galaxies and Cosmology, School of Astronomy and Space Science, University of Science and Technology of China, Hefei, Anhui 230026, China}

\author{Chao Chen}
\email{iascchao@ust.hk}
\affiliation{Jockey Club Institute for Advanced Study, The Hong Kong University of Science and Technology, \\Clear Water Bay, Kowloon, Hong Kong, China}

\author{Qianhang Ding}
\email{qdingab@connect.ust.hk}
\affiliation{Department of Physics, The Hong Kong University of Science and Technology, \\Clear Water Bay, Kowloon, Hong Kong, China}
\affiliation{Jockey Club Institute for Advanced Study, The Hong Kong University of Science and Technology, \\Clear Water Bay, Kowloon, Hong Kong, China}

\author{Yi Wang}
\email{phyw@ust.hk}
\affiliation{Department of Physics, The Hong Kong University of Science and Technology, \\Clear Water Bay, Kowloon, Hong Kong, China}
\affiliation{Jockey Club Institute for Advanced Study, The Hong Kong University of Science and Technology, \\Clear Water Bay, Kowloon, Hong Kong, China}

\begin{abstract}
	In this article we study a hypothetical possibility of tracking the evolution of our Universe by introducing a series of the so-called {\it standard timers}. 
	Any unstable primordial relics generated in the very early Universe may serve as the standard timers, as they can evolve through the whole cosmological background until their end while their certain time-varying properties could be a possible timer by recording the amount of physical time elapsed since the very early moments. 
	Accordingly, if one could observe these quantities at different redshifts, then a redshift-time relation of the cosmic history can be attained. 
	To illustrate such a hypothetical possibility, we consider the primordial black hole bubbles as a concrete example and analyze the mass function inside a redshifted bubble by investigating the inverse problem of Hawking radiation. 
	To complete the analyses theoretically, the mass distribution can serve as a calibration of the standard timers.  
\end{abstract}

\maketitle

\section{Introduction} \label{sec:intro}

The cosmology of $\Lambda$-cold dark matter ($\Lambda \text{CDM}$) is widely acknowledged as the most successful theory in depicting the evolution of the Universe, which has been examined in various high-precision observations, namely, the cosmic microwave background (CMB) \cite{PhysRev.74.505.2, PhysRev.74.1737, PhysRev.75.1089}, large scale structure (LSS) \cite{Peebles:1982ff}, and the acceleration at late times \cite{SupernovaCosmologyProject:1998vns, SupernovaSearchTeam:1998fmf}, etc. However, there remain puzzles in the Universe, such as the recent focuses on the Hubble tension \cite{Riess:2021jrx, Riess:2016jrr, Planck:2016kqe} and $\sigma_8$ tension \cite{Planck:2018vyg}, which hint on the existence of unknowns beyond the $\Lambda \text{CDM}$.

Accompanied with the dramatic developments of observational technologies, various approaches of astronomical measurements have been proposed to study the evolution of the Universe. Type Ia supernovae (SNe) produces consistent peak luminosity after calibration, of which the distances can measure by comparing their absolute and apparent magnitudes, and hence they can serve as a standard candle \cite{1969PASP...81..707F}. Baryon acoustic oscillations (BAO) determine the fixed scale that the sound wave can travel before the recombination, through probing the BAO scale at different redshifts, and then the cosmological models can be effectively constrained, and accordingly, BAO can work as a standard ruler \cite{Beutler:2011hx, Ross:2014qpa}. Gravitational waves (GWs) and electromagnetic (EM) waves from binary compact objects and EM counterpart provide the calibration between luminosity distance and redshift, which makes GWs standard sirens in constraining cosmological parameters \cite{LIGOScientific:2017vwq}. 
While the aforementioned methods can constrain cosmological models, the distinct physics behind them would bring difficulties in excluding the unrecognized systematic uncertainties and may still cause the tension for cosmological parameters. 
Moreover, most observational signals lie within certain redshift windows. Standard candle via SNe covers redshift window $0.01 < z < 2.3$ \cite{Pan-STARRS1:2017jku}. Standard ruler via BAO by observing galaxies can be applied mostly on $0.106 \lesssim z \lesssim 0.61$ \cite{Beutler:2011hx, Ross:2014qpa}, the signals from $\text{Ly}\alpha$ forest of high-redshift quasars \cite{BOSS:2014hwf} and CMB \cite{Planck:2018vyg} provide the acoustic scales at $z = 2.34$ and $1100$, respectively. Standard siren observations based on aLIGO are only allowed within $z < 1$ \cite{Bai:2018shq, Ding:2020ykt}.

In contrast, unstable primordial relics can be used to track the whole evolution history of the Universe. If these relics are unstable, the decay of the relics tracks the physical time of the Universe, just as standard timers distributed in the Universe. If we observe distant standard timers, a redshift-time relation is obtained, which can be used to constrain cosmological models.

In this paper, we discuss an explicit model of the standard timer, i.e.,  primordial stellar bubbles consisting of primordial black holes (PBHs) in tracking the evolution of the Universe. The PBH stellar bubbles can be generated from some primordial physics, such as multi-stream inflation \cite{Li:2009sp, Ding:2019tjk}, which causes a number of PBHs clustering in specific regions. Its primordial origin produces the same PBH mass functions inside PBH stellar bubbles. Since the observed signals from PBH bubbles are determined by the internal mass function evolution and external cosmic expansion, the analysis of the observed signals can construct the one-to-one correspondence between the PBH bubble internal mass functions and the evolution of the Universe. In the pioneer paper \cite{Cai:2021zxo}, we studied the detectability of such PBH bubbles with the current and the near future observations, and presented the detailed analyses of the possible observational signals from such a stellar bubble, including the EM and GW signals, coming from the light PBHs' Hawking radiations and binary mergers of PBHs inside a bubble. Limited by the current resolution of gamma-ray detectors, these PBH bubbles can be regarded as the exotic celestial objects. After having used the point-source differential sensitivity in the 10-year observation of Fermi LAT for a high Galactic latitude (around the north Celestial pole) source \cite{Atwood:2009ez}, we showed that the lowest detectable bubble mass is $M_\text{bub} \sim 10^{32}$ g with the peak mass $10^{15}$ g of the lognormal distribution of PBH masses inside the bubble. Also, we found that massive PBH binaries with $M_\text{pk} \sim 10^{34} - 10^{38}$ g and $M_{\mathrm{bub}} \sim 10^{45} - 10^{48}$g can produce detectable GWs within in the frequency band of LISA and BBO. Hence, EM and GW signals are complementary for light and heavy PBH bubbles. Since PBH bubbles can be regarded as point-sources for the current EM observations (like the Fermi LAT experiment), the redshift information of PBH bubbles is thus imprinted in their EM signals. As we will show below, one can infer the redshift information of PBH bubbles through the inverse process, which is essential to the determination of the time evolution of our Universe. In the following context, we focus on the EM signals coming from PBH bubbles and the role that PBH bubbles can play in , hence we call it the {\it standard timer}. 

This paper is organized as follows. We present a brief introduction of PBH bubbles in Sec.~\ref{sec:PBHBubble}; In Sec.~\ref{sec:Timer}, we study the inverse problem for Hawking radiation to extract the mass function of a PBH bubble from EM observations. Then, we describe the approach of calibrating the redshifts of PBH bubbles from the inverse results and construct the redshift-time calibration through the standard timer. In Sec.~\ref{sec:TestLCDM}, we apply the redshift-time calibration from the standard timer to test the standard cosmological model. The conclusion and discussions are given in Sec.~\ref{sec:concl}.

\section{PBH Bubbles} \label{sec:PBHBubble}

PBHs have been attracting a lot interest for decades. They can be formed from overdense regions that possibly exist in the early Universe \cite{Hawking:1971ei, Carr:1974nx, Carr:1975qj}. In contrast to the regular stellar-origin BHs, PBHs in general possess a quite wide range of masses, from tens of micrograms to millions of solar masses. Hence, PBHs can be related to various cosmological and astrophysical phenomena. For example, for the heavy PBHs ($M \geq 10^{15}$ g), they could be a reasonable candidate for dark matter (DM), however, the open windows for PBHs to be the whole DM have been tightly limited nowadays \cite{Carr:2020xqk}. While, for the very light PBHs ($M \leq 10^{15}$ g), their Hawking radiation can be strong \cite{Hawking:1974rv, Hawking:1974sw} and have already evaporated, the Hawking-radiated particles may influence physical processes in the early Universe, like Big Bang Nucleosynthesis \cite{Kohri:1999ex, Carr:2009jm, Luo:2020dlg}. Meanwhile, with the dramatic developments of GW experiments, a lot of attentions focus on the GW signals coming from PBHs associated with the binary mergers \cite{Bird:2016dcv, Sasaki:2016jop, Ali-Haimoud:2017rtz, Chen:2018czv, Ding:2020ykt, Ding:2019tjk} and scalar perturbations \cite{Saito:2008jc, Kohri:2018awv, Cai:2019cdl, Cai:2019jah, Inomata:2018epa, Inomata:2019yww, Fu:2019ttf, Inomata:2020cck, Cai:2021wzd, Domenech:2020kqm, Peng:2021zon, Pi:2019ihn, Cai:2018dig, Bartolo:2018rku, Zhou:2020kkf, Chen:2022qec, Fu:2022ssq, Chen:2023lou, Pi:2022ysn, Pi:2022zxs}. In summary, PBHs can be a promising needle to detect physics in the early and late Universe.

In usual cases, PBHs are regarded as individual isolated objects in the Universe, however, there are many scenarios for clustering of PBHs at some certain scales. If the sizes of these regions are small enough, say, smaller than the resolutions of current telescopes, they behave as exotic celestial objects, i.e., the PBH bubbles \cite{Ding:2019tjk, Cai:2021zxo}. Generally speaking, these stellar bubbles can be generated from some new-physics phenomena that might have occurred in the primordial Universe, such as, quantum tunnelings during or after inflation \cite{Coleman:1980aw, Zhou:2020stj}, multi-stream inflation \cite{Li:2009sp, Duplessis:2012nb, Cai:2021hik}, inhomogeneous baryogenesis \cite{Cohen:1997ac}, etc. In these cases, before the bubble-wall tension vanishes, the field values are different between inside and outside of the bubble. Such difference can result in different local physics inside the bubble (for PBH cases, see \cite{Belotsky:2018wph, Ding:2019tjk} for details), namely the production rate of the exotic species of matter. 

In general, the abundance of bubbles is determined by the probability of the tunneling (for phase transition) or bifurcation (for multi-stream inflation). And the size of the bubbles is determined by the comoving scale at which tunneling or bifurcation happened. For example, in the multi-stream inflation model, the radius of the bubble is similarly $R_b = R_0\exp(-N_b)$, where $R_0$ is the radius of the current observable Universe and $N_b$ is interpreted as the e-folding number between the beginning of the observable inflation to the bifurcation. Since the bifurcated path eventually merges, the tension of the bubble wall vanishes automatically. The number density of the bubble $n_b$ is determined by the shape of the multi-field potential, and the amplitude of the isocurvature fluctuation during inflation. We should keep in mind that the local abundance of PBHs inside a bubble $f_\text{local}\equiv \Omega_\text{PBH}/\Omega_\text{tot}$ is naturally enhanced by the probability of formation of this bubble, such as the bifurcation probability $f_\text{bifu} < 1$, see the Fig.1 in Ref. \cite{Ding:2019tjk}. We can write $f_\text{local} = f_\text{glob} f_\text{bifu}^{-1}$, where $f_\text{glob} < 1$ is the normal definition of the abundance of PBHs over the whole observable Universe \cite{Carr:2020xqk}. So, it is expected that the enhanced observable signals coming from these PBH bubbles, like the detectable GW and EM signals with the current observations \cite{Ding:2019tjk, Cai:2021zxo}. Notably, \cite{Cai:2021zxo} shows that even for PBH bubbles at quite high reshifts $z \sim 10^3$, their EM signals are possible to be detected by FermiLAT experiment. So that in this sense, PBH bubbles are able to track the history of the Universe over a long period.

\section{PBH Bubbles as a Standard Timer} \label{sec:Timer}

The same primordial origins of PBH bubbles are expected to be formed at a nearly single moment with the same initial mass functions \cite{Ding:2019tjk}. Then, PBHs inside a bubble would evaporate through Hawking radiations, which results in the deformation of their mass function insides the bubble. We notice that there are two key quantities encoded in this process. One is the cosmic expansion which causes the redshift of the observed photon energies, the other is the physical evolution time that hides in the deformed mass function. The cosmological redshift depends on the evolution of the Universe, while the physical evolution time is determined by Hawking evaporation and does not rely on the cosmological redshift. Hence, the independent channels of cosmological redshifts and physical times indicate that PBH bubbles can calibrate the redshift-time relation, which works as a standard timer of the Universe.

In order to extract cosmological redshifts and physical times from the EM signals of PBH bubbles, two problems need to be resolved. One is that the redshift cannot be obtained directly from the observed photon energy due to the unknown emission energy of photon. However, the observed photon flux spectrum is a result from the interplay of the Hawking radiations emitted over the PBH mass function and the cosmological redshifts, which enable us to obtain the redshifted PBH mass function from the observed photon flux spectrum and thus the redshift is encoded in it. This method is discussed in this section and the details presented in Appendix \ref{inverseBlackbody} and \ref{inverseHawking}. There are some other methods for helping estimate the redshift of PBH bubbles such as the redshifts from the host galaxies or neighbour astrophysical objects \cite{Bloom:2000nj}. The other problem is to determine the physical evolution times of PBHs in lack of their primordial mass function. This issue can be addressed when the redshift is extracted from the observed photon flux spectrum, the physical PBH mass function can be extracted from the redshifted mass function, then physical evolution time between two PBHs bubbles can be obtained by comparing their physical PBH mass functions. In this section, we will introduce the method of extracting the redshifted mass function of PBHs from observed photon flux spectrum, then obtain the redshift from the redshifted mass function, and construct the standard timer array in the Universe.

\subsection{Inverse Problem for Hawking Radiation}

The Hawking radiation spectrum $R(E)$ emitted from a PBH bubble follows
\begin{align}\label{eq:inverHawking}
 R(E) = \int_0^{\infty} H(E, M) n(M)dM ~,
\end{align}
where $H(E, M)$ is Hawking radiation kernel and $n(M)$ is the PBH mass function. In order to study the evolution of PBH mass function $n(M)$, extracting the PBH mass function from the observed emission rate spectrum is a key step, and we call it the inverse problem for Hawking radiation. The similar problem has already been studied for decades, i.e., determining the temperature distribution on the thermal radiator for a given radiation power spectrum, which is known as the inverse black-body radiation problem \cite{1142844}. There are several inversion methods that have been proposed to solve the inverse black-body radiation problem, such as iteration method and M\"obius inversion transformation \cite{1142844, PhysRevLett.64.1193} (see Appendix~\ref{inverseBlackbody} for more details). 

In studying the inverse Hawking radiation problem, the properties of the Hawking radiation kernel $H(E, M)$ and emission rate spectrum $R(E)$ are crucial in analysis. The Hawking radiation kernel can be divided into two parts, the primary emission $H_{p}(E, M)$ (the direct Hawking emission) and the secondary emission $H_{s}(E,M)$ (from the decay of gauge bosons or heavy leptons and the hadrons produced by the fragmentation of primary quarks and gluons \cite{MacGibbon:1990zk}), and we write
\begin{align}
 H(E,M) = H_{p}(E,M) + H_{s}(E,M) ~.
\end{align}
Since the gamma ray observations are more sensitive in the high-energy range and the primary emission dominates such Hawking radiation \cite{Carr:2020gox}, we focus on the primary emission in the inverse Hawking radiation problem. On the other hand, the secondary emission involves the fragmentation and hadronization of quark and gluon jets, which lacks the precise analytic expressions, especially in the low-energy range \cite{MacGibbon:1991tj}. Moreover, one can in principle involve the secondary contribution by virtue of the numeric calculations, we leave this to the future study. In practice, the emission rate spectrum $R(E)$ comes from the high-energy gamma ray observation, which means $R(E)$ is in the form of a data array instead of an analytic expression. As a result, we should apply discretization on the Eq.~\eqref{eq:inverHawking}, which is in form of
\begin{align}
 R(E_i) \simeq \sum_{j = 1}^{N} w_j H(E_i, M_j)n(M_j) ~.
\end{align}
Here, $w_j$ is the weight of the quadrature formula, $N$ is the number of discretization nodes for the mass function $n(M)$. Then, Eq.~\eqref{eq:inverHawking} is transformed to a linear algebraic equation, we can apply the method for the least squares problem (see Ref. \cite{lawson1995solving} for more details) to resolve the mass function vector as follows,
\begin{align}\label{eq:leastsquare}
 ||\textbf{R} - \textbf{Kn}||_2^2 = \text{minimum} ~.
\end{align}
Here, $\textbf{R}$ and $\textbf{n}$ are the emission rate vector and mass function vector, respectively. $\textbf{K}$ is the kernel matrix, whose element is $\textbf{K}_{ij} = w_j H(E_i, M_j)$. Actually, due to the observational error existed in the emission rate vector $\textbf{R}$, there exist a number of potential vectors $\textbf{n}$ for the Eq.~\eqref{eq:leastsquare}, and the regularization methods \cite{PROVENCHER1982213} can be applied to determine the physical mass function vector. We also show the formal method for the inverse Hawking radiation problem in Appendix~\ref{inverseHawking}.

In the observed photon flux from a PBH bubble, the photon energy is redshifted by the cosmic expansion. The photon flux can be expressed as
\begin{align}\label{eq:flux}
 F(E; z) = \frac{L(E(1+z); z)}{4 \pi d_L^2(z)} ~.
\end{align}
Here, $L(E; z)$ is the intrinsic luminosity emitted from the PBH bubble at redshift $z$ which can be calculated as $L(E; z) \simeq E^2 R(E; z) V(z)$, where $V(z)$ is the volume of the PBH bubble.
Then, Eq.~\eqref{eq:flux} can be written as
\begin{align}\label{eq:flux_detail}
	F(E; z) = \frac{L(E(1+z); z)}{4 \pi d_L^2(z)} \simeq \frac{(1+z)^2 E^2 V(z)}{4 \pi d_L^2(z)} \int_{0}^{\infty}  H_p(E(1 + z),M) n(M; z) dM ~.
\end{align}
There is $1 + z$ term appearing in the Hawking kernel, and the unknown redshift causes the uncertainty in determining the PBH mass function $n(M; z)$ in the inverse problem. In general, we can randomly choose the redshift in the Hawking kernel until the proper mass function returns, however, the error in introducing unknown redshift may increase the error in mass function. Consequently, we can transform $1 + z$ term from the Hawking kernel to the mass function, which results in the redshift uncertainty in the mass function. In the primary emission of the Hawking radiation $H_p(E, M)$, the $E$ and $M$ terms in $H_p(E,M)$ are symmetry, see Eq. \eqref{greyfactor}, so we have
\begin{align}
    H_p(E(1 + z),M) = H_p(E,M(1 + z)) ~.
\end{align}
Then, Eq.~\eqref{eq:flux_detail} can be written as
\begin{align}\label{eq:flux_to_mass}
	F(E; z) \simeq& \frac{(1+z)^2 E^2 V}{4 \pi d_L^2(z)} \int_{0}^{\infty}  H_p(E, M(1+z)) n(M; z) dM \nn
	\\=& \frac{(1+z)^2 E^2 V}{4 \pi d_L^2(z)} \int_{0}^{\infty}  H_p(E, M') n(\frac{M'}{1+z}; z) \frac{1}{1+z} dM' ~.
\end{align}
Here, $M' \equiv M(1+z)$. Then we have the equation,
\begin{align}
	\frac{4 \pi F(E; z)}{E^2} \simeq \int_{0}^{\infty}  H_p(E, M) n(\frac{M}{1+z}; z) \frac{(1+z) V}{d_L^2(z)} dM ~.
\end{align}
Following the method for the inverse Hawking radiation problem in Eq.~\eqref{eq:leastsquare}, we can obtain the redshifted mass function $f(M; z)$,
\begin{align}
    f(M; z) = n(\frac{M}{1+z}; z) \frac{(1+z) V}{d_L^2(z)} ~.
\end{align}

\subsection{Redshift Calibration}

\begin{figure}[ht]
\centering
\includegraphics[width=8cm]{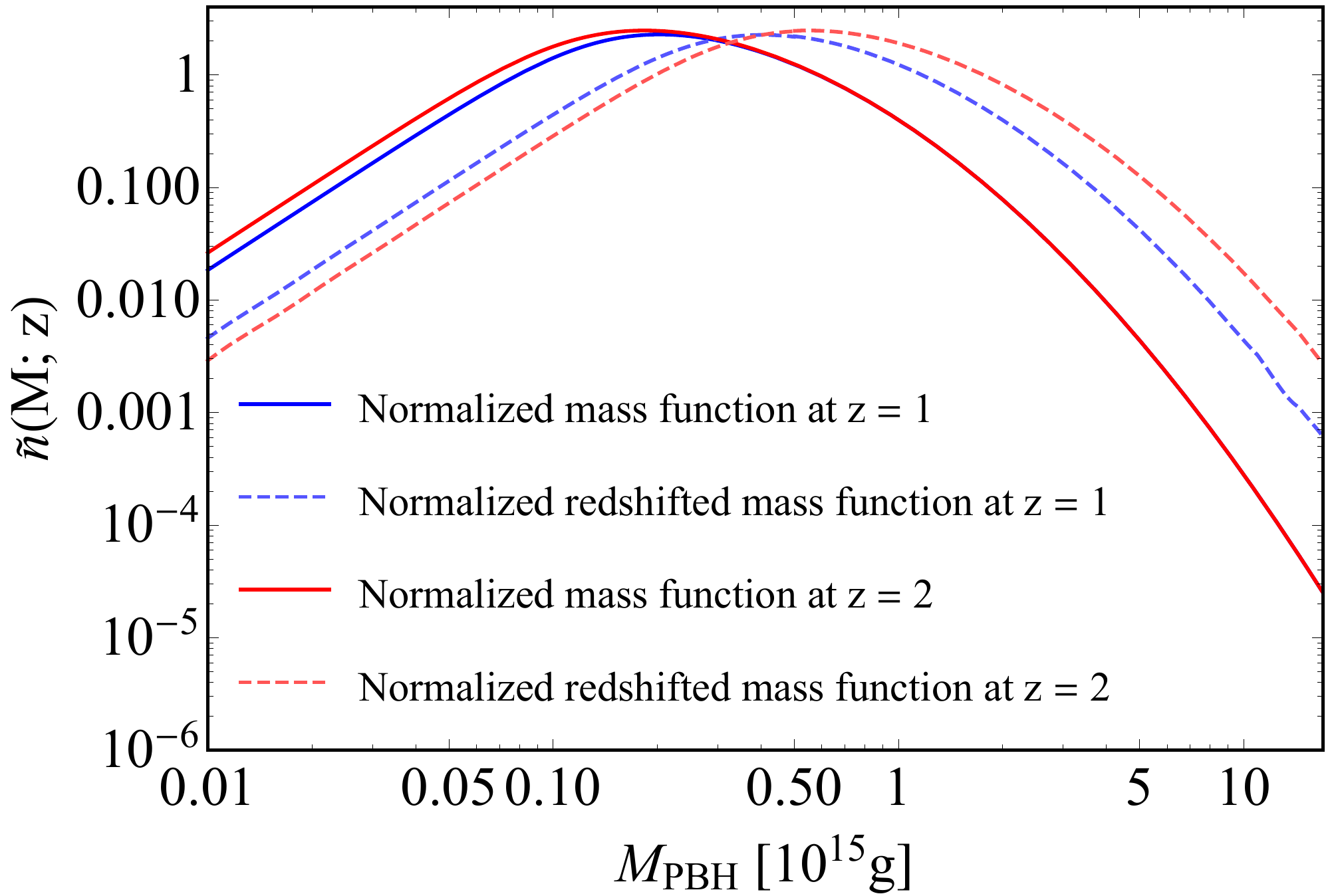}
\caption{The normalized lognormal mass function with $M_{pk} = 10^{15}\mathrm{g}$ and $\sigma = 1$. The blue and red solid curves are the mass functions evolving from their formation to the redshifts $z = 1$ and $z = 2$, respectively. The blue and red dashed curves denote the normalized redshifted mass functions obtained from the inversion of Hawking radiation emitted at redshifts $z = 1$ and $z = 2$, respectively.}
\label{fig:redmass}
\end{figure}

In the redshifted mass function $f(M; z)$, the redshift $z$ is unknown, which is essential for the standard timer. Considering two PBH bubbles with redshifts $z_1$ and $z_2$, we get the redshifted mass functions
\begin{align}\label{eq:two_redmass}
 f_1(M; z_1) = n(\frac{M}{1+z_1}; z_1) \frac{(1+z_1) }{d_L^2(z_1)} V_1 ~,~~~	
 f_2(M; z_2) = n(\frac{M}{1+z_2}; z_2) \frac{(1+z_2) }{d_L^2(z_2)} V_2 ~.
\end{align}
Here, a factor $1/(1 + z)$ appears in the argument of mass function $n(M)$ which transforms the argument $M \to M/(1 + z)$ in $n(M)$. By comparing the locations of the large-mass tails of these two normalized redshifted mass functions, where the Hawking radiation effect is negligible and the large-mass tails remain nearly unchanged (see the blue and red solid curves in Fig.~\ref{fig:redmass}), we obtain the redshift ratio $\eta$:
\begin{align}
	\eta \equiv \frac{1 + z_2}{1 + z_1}~.
\end{align}
For an illustration, we take a normalized lognormal mass function [see Eq. \eqref{lognormal}] at redshifts $z = 1$ and $z = 2$ shown in Fig.~\ref{fig:redmass}, which gives the $\eta \approx 1.5005$ with the error $0.03\%$.

For a given redshift $z_1$, $z_2 = \eta (1 + z_1) - 1$, we can reconstruct the normalized PBH mass function at $z_1$ and $z_2$ as follows,
\begin{align}
	\tilde{n}(M; z_1) \equiv \tilde{f}_1(M(1 + z_1); z_1)~,~~~
	\tilde{n}(M; z_2) \equiv \tilde{f}_2(M(1 + z_2); z_2)~.
\end{align}
Here, $\tilde{n}$ and $\tilde{f}$ denote the normalized comoving mass functions. Then, the physical evaporation times can be extracted from $\tilde{n}(M; z_1)$ and $\tilde{n}(M; z_2)$. Suppose that the physical evaporation time between $z_2$ and $z_1$ is $t_m$, we write the mass evaporation relation [see Eq. \eqref{mass_time_evo}] as
\begin{align}\label{eq:mass_rel}
	M^3_{z_2} = M^3_{z_1} + \delta^3(t_m)~ ,
\end{align}
where $\delta(t_m)$ is a term corresponding to the evaporated mass during the period $z_2$ to $z_1$. Then, we construct the relation between $\tilde{n}(M; z_1)$ and $\tilde{n}(M; z_2)$ as
\begin{align}\label{eq:massdis_rel}
	\tilde{n}(M_{z_1}; z_1) = \frac{d\tilde{N}}{dM_{z_1}} = \frac{d\tilde{N}}{dM_{z_2}}\frac{dM_{z_2}}{dM_{z_1}} = \tilde{n}(M_{z_2}; z_2)\frac{dM_{z_2}}{dM_{z_1}}~.
\end{align}
Applying Eq.~\eqref{eq:mass_rel} into Eq.~\eqref{eq:massdis_rel}, we obtain
\begin{align}
	\tilde{n}(M_{z_1}; z_1) = \tilde{n}((M^3_{z_1} + \delta^3(t_m))^{1/3}; z_2)\frac{M^2_{z_1}}{(M^3_{z_1} + \delta^3(t_m))^{2/3}}~.
\end{align}
In the low-mass approximation $M_{z_1} \ll \delta(t_m)$ which holds for the small values of $M_{z_1}$, since the most of PBH's masses have been evaporated for small $M_{z_1}$. Thus, we obtain,
\begin{align}\label{eq:lowmass}
	\log \tilde{n}(M_{z_1}; z_1) \simeq \log \frac{\tilde{n}(\delta(t_m); z_2)}{\delta^2(t_m)} + 2 \log M_{z_1}~.
\end{align}
Here, the term $\tilde{n}(\delta(t_m); z_2)/\delta^2(t_m)$ can be extracted from $\tilde{n}(M_{z_1}; z_1)$ in low-mass approximation, which gives the $\delta(t_m)$ from $\tilde{n}(M_{z_2}; z_2)$. The physical evaporation time $t_m$ is thus extracted from $\delta(t_m)$. Impressively, the low-mass tail on the logarithmic scale possesses the universal slope $d \log n(M)/d \log M \simeq 2$, which is independent on the initial PBHs' mass function, arising from the properties of mass evaporation \cite{Carr:2016hva}, which is also confirmed by our following analysis, see Fig. \ref{fig:MfToM}.

In order to determine the redshift $z_1$, we need to compare the physical evaporation time $t_m$ with the cosmic evolution time $t_z$ between $z_1$ and $z_2$. The proper $z_1$ should be chosen such that the physical evaporation time is same as the cosmic evolution time as follows:
\begin{align}
	t_m = t_z~,
\end{align}
where $t_z$ is assumed to follow the standard cosmology evolution, which is calculated as
\begin{align}
	t_z = \int_{z_1}^{z_2} \frac{dz}{H(z)(1 + z)}~.
\end{align}
Then the redshift of one PBH bubble is calibrated.

\begin{figure}[ht]
\centering
\includegraphics[width=8cm]{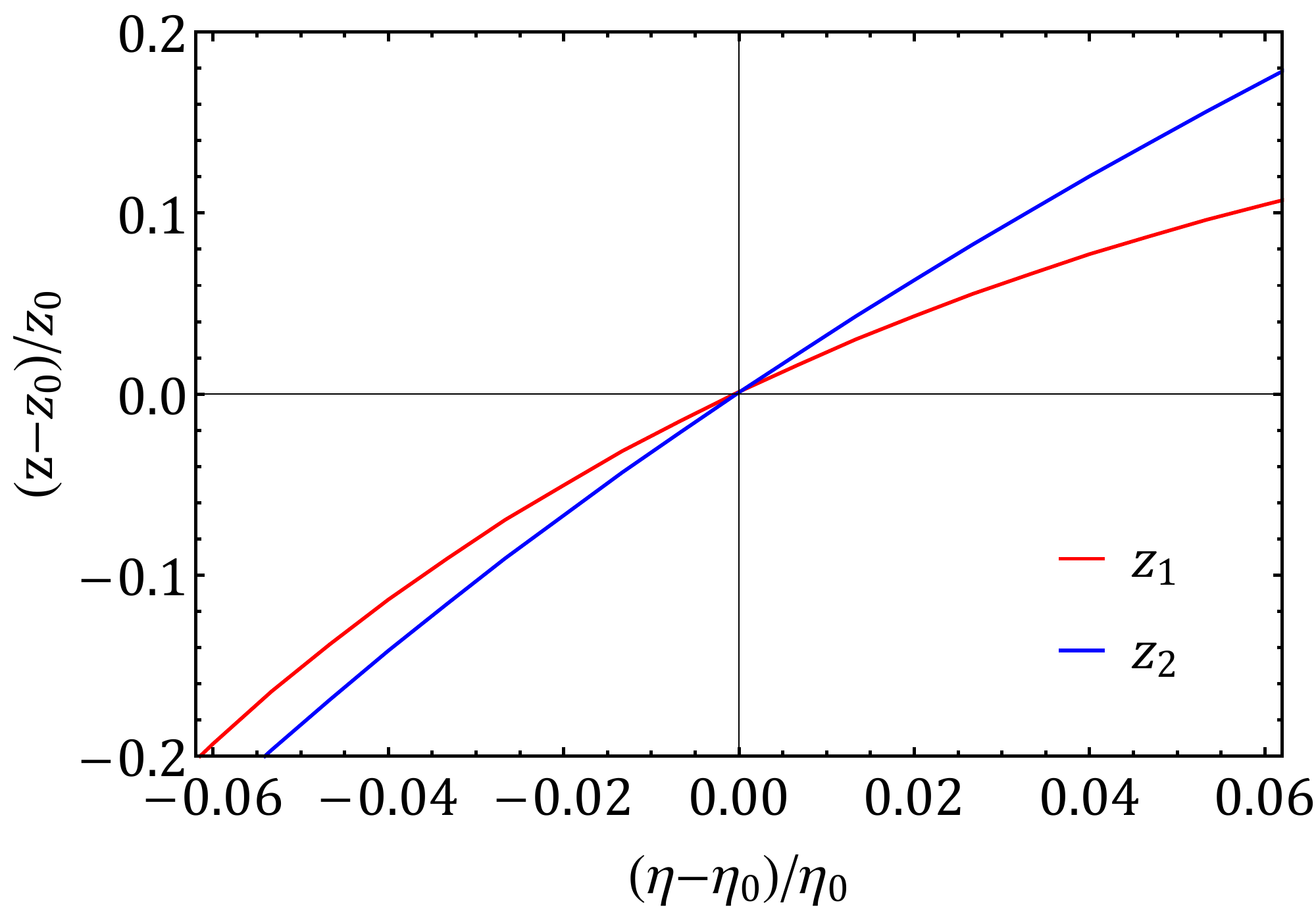}
\caption{The error in redshift calibration in terms of the error in $\eta$ measurement. The redshift calibration is based on two PBH bubbles at different redshifts. We choose the lognormal mass function with $M_{pk} = 10^{15}\mathrm{g}$ and $\sigma = 1$ inside PBH bubbles and take $z_1 = 1$ and $z_2 = 2$ as an example in calculation.}
\label{fig:red_error}
\end{figure}

The accuracy of the calibrated redshift $z$ depends on the redshift ratio $\eta$. A practical method for redshift calibration should be stable, i.e., the error in calibrated redshift can not be much larger than the error in the redshift ratio. We take the PBH bubble with the same parameters setting in Fig.~\ref{fig:redmass} as an example, see Fig.~\ref{fig:red_error}, which gives the calibrated redshifts $z_1 \simeq 1.002$ and $z_2 \simeq 2.004$.

\subsection{Standard Timer via Redshift-Time Calibration}

\begin{figure}[ht]
\centering
\includegraphics[width=8cm]{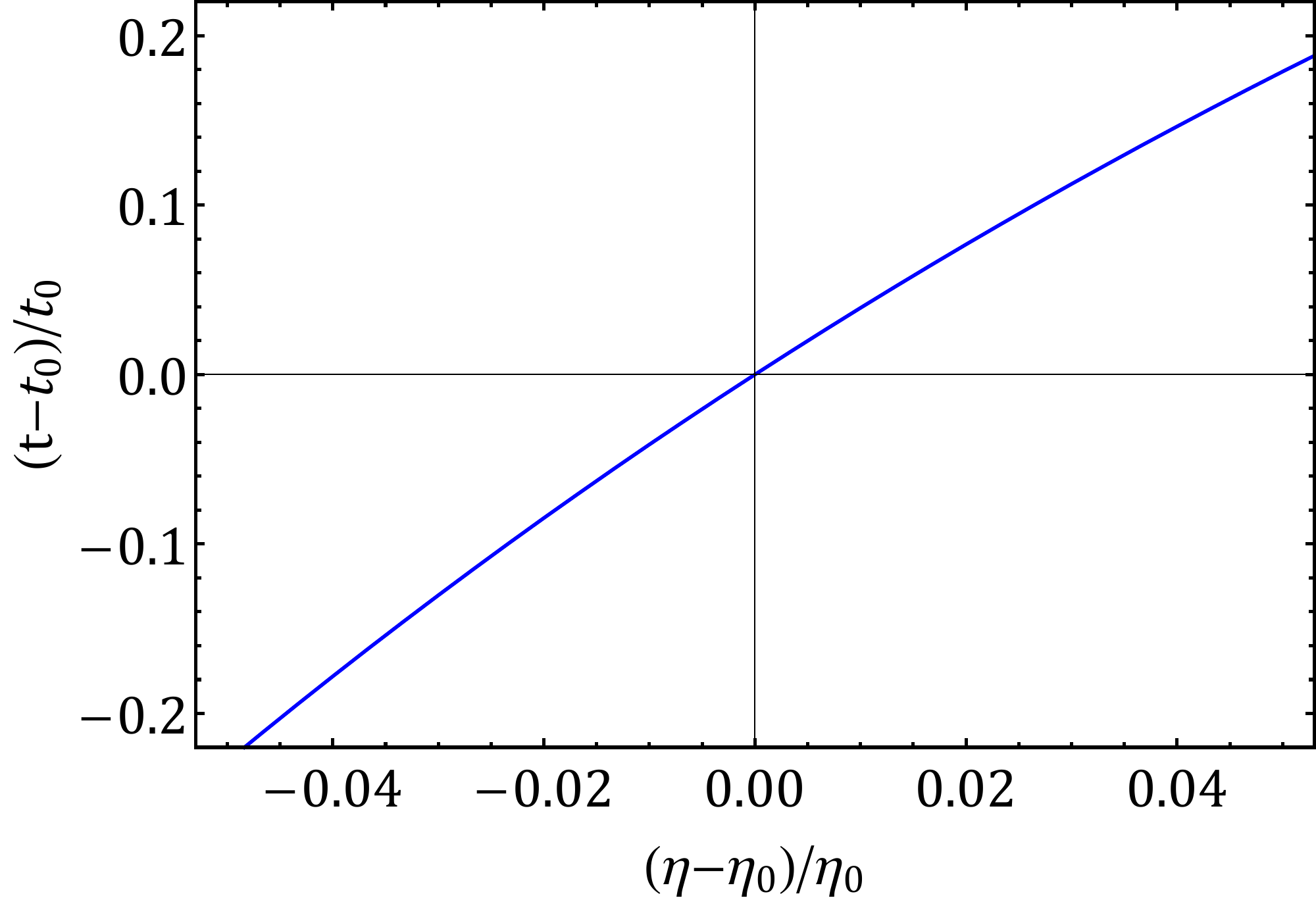}
\caption{The error in the time calibration in terms of the error in $\eta$ measurement. We choose the lognormal mass function with $M_{pk} = 10^{15}\mathrm{g}$ and $\sigma = 1$ inside a PBH bubble and the redshift calibrator $z = 1$ as an example in calculation.}
\label{fig:time_error}
\end{figure}

As we have discussed above, the redshift of the PBH bubble is calibrated, which can work as a redshift calibrator for other PBH bubbles. The redshift ratio $\eta$ between a PBH bubble and the redshift calibrator can be determined by matching the redshifted mass functions. As a result, the redshift of a PBH bubble $z_{\mathrm{PBH}}$ is given by
\begin{align}\label{eq:pbhred}
	z_{\mathrm{PBH}} = (1 + z_0) \eta - 1~,
\end{align}
where $z_0$ is the redshift of the redshift calibrator. Then physical time interval $\Delta t$ between $z_0$ and $z_\mathrm{PBH}$ should equal to the physical evolution time between the redshift calibrator and the PBH bubble in Eq.~\eqref{eq:lowmass},
\begin{align}
	\Delta t = t_m~.
\end{align}
The accuracy of the calibrated physical time interval $\Delta t$ depends on the redshift ratio $\eta$. In order to calibrate the $\Delta t$ with high precision, we need to estimate the error in the calibrated time versus the error in redshift ratio, see Fig.~\ref{fig:time_error}, which produces the error in time calibration is around $0.1 \%$, while the error in $\eta$ is obtained around $0.03\%$ and the redshift calibrator is set to $z = 1$.

Consequently, the redshift-time calibration $(z_0, z_\mathrm{PBH}, \Delta t)$ is obtained. After detecting a bunch of PBH bubbles in the Universe, a redshift-time calibration array can be constructed cover from the local Universe to the primordial Universe, which helps us constrain the cosmological models as the standard timer in the Universe.

\subsection{The Time-Dependent Mass Function} \label{sec:MassFunc}

In the above discussions, we have presented the formal framework of a standard timer made of PBH bubbles. Here, we derive the approximated analytic formulas for the time evolution of PBHs' mass function due to their Hawking radiation, which enables us to extract the physical evolution time of PBH bubbles by using the methods studied in the above part.

We adopt the standard Hawking evaporation picture \cite{MacGibbon:1990zk}: a black hole would directly radiate fundamental standard model particles whose de Broglie wavelengths are of the order of black hole size. Hence, PBHs would radiate heavier particles successively with time, as shown in the Table \ref{tab:hawking}. 
\begin{table}[h] 
	\caption{The approximate values of $f(M)$.}
	\centering
	\begin{tabular}{||c | c ||}
		\hline
		PBH mass/ M (g) & f(M) \\
		\hline
		$\gg 10^{17}$ & $2 \times 0.06 (\text{photon}) +6 \times 0.147 (\text{neutrino}) + 2 \times 0.07 (\text{graviton}) = 1.142$ \\
		\hline
		$10^{15} \sim 10^{17}$ & $1.142 + 4 \times 0.142 (\text{electron}) = 1.71$  \\ 
		\hline
		$10^{14} \sim 10^{15}$ & $1.71 + 4 \times 0.142 (\text{muon}) = 2.278$ \\
		\hline
		$10^{12} \sim 10^{14}$ & $2.278 + 3 \times 12 \times 0.147 (\text{u,d,s quarks}) + 16 \times 0.06 (\text{gluon}) = 8.53$  \\
		\hline
		$\leq 10^{12}$ & $\text{(b,c,t quarks) + (tau) + (gluon) + (W, Z, Higgs)} \simeq 15$  \\
		\hline
	\end{tabular}
\label{tab:hawking}
\end{table}
According to the Hawking radiation and energy conservation, we derive the mass-loss rate for a single PBH as follows,
\be
\frac{d M(t)}{d t} = - \frac{\phi[M(t)]}{3 M(t)^2} ~,
\ee
where $\phi(M)$ measures the number of emitted particle degrees of freedom for a PBH with mass $M$ and the factor ``$3$" is included for convenience. The relativistic contributions to $\phi(M)$ per degree of particle freedom are \cite{MacGibbon:1990zk}
\bl
& f_{s=0} = 0.267,~~ f_{s=1} = 0.060,~~f_{s=3/2} = 0.020, ~~f_{s=2} = 0.007, \nn
\\& f_{s=1/2} = 0.147 ~(\text{neutral}),~~ f_{s=1/2} = 0.142 ~(\text{charge} \pm e) ~,
\el
where we introduced the dimensionless parameter $f \equiv \phi/(5.34 \times 10^{25} ~ \text{g}^3 ~ \text{s}^{-1})$. Hence, the PBH's mass at time $t$ can be written as the following formula:
\be \label{mass_time_evo}
M(t)^3 = M_f^3 - \int_{t_f}^{t} \phi[M(\tilde{t})] d\tilde{t} ~,
\ee
where $M_f$ is the PBH mass at the formation epoch $t_f$. According to this formula, we can also derive the lifetime of a single PBH with an arbitrary initial mass. 

In order to make more precise analysis for the light PBHs which dominates EM signals from PBH bubbles, we extended the approximation for $\phi[M(t)]$ used in Ref. \cite{Carr:2016hva}: the small mass region is further separated, i.e., $M \leq M_g$, $M_g$ is the mass scale that the W, Z, Higgs bosons are directly emitted from such PBHs, while the large mass is also extended. We write:
\be \label{phi_appro}
\phi \simeq
\l\{
\begin{aligned}
	& \kappa \phi_*, ~~~&& M \geq 10 M_*
	\\& \phi_*,      && M_q \leq M \leq 10 M_*
	\\& \alpha \phi_*, && M_g \leq M \leq M_q
	\\& \omega \phi_*, && M \leq M_g
\end{aligned}
\r.
~,
\ee
where $\kappa \simeq 0.5$, $\alpha \simeq 4$ and $\omega \simeq 8$ to sufficient precision, $M_*$ denote those PBHs' masses which are completing the evaporation at present, see the discussion below. When the PBH's mass goes down $M_q$, the secondary emission would be triggered. Note that, in order to calculate the time-evolution of PBHs' masses, there are several characteristic mass scales of great interest. 
\begin{itemize}
	
	\item $M_a$:
	
\end{itemize}

First, let us see how long would a PBH with formation mass $M_f > 10 M_*$ fall into $10 M_* \simeq 5.1 \times 10^{15}$ g:
\be
t_*(M_f) \simeq \frac{M_f^3 - (10 M_*)^3}{\kappa \phi_*}
\equiv
t_0 \frac{M_f^3 - (10 M_*)^3}{ \kappa \bar{M}_*^3} ~,
\ee
where we also define
\be
\bar{M}_* \equiv (\phi_* t_0)^{1/3} \simeq 5.07 \times 10^{14} \l( \frac{f_*}{1.9} \r)^{1/3} \text{g}
\ee
is the mass of a PBH currently evaporating if one neglects secondary emission once $M(t)$ falls below $M_q$, and we have used the approximation $t_f \simeq 0$, and $t_0 = 13.8$ Gyr is the lifetime of Universe. The value of $f_*$ is around $1.9$ that can be derived by counting the quantum dofs of Hawking-radiated particles. Consider a formation mass $M_f = M_a > 10 M_*$, whose corresponding present mass is exact $10 M_*$, we can derive
\be
M_a = [ (10 M_*)^3 + \kappa \bar{M}_*^3 ]^{1/3}
\simeq 5.15 \times 10^{15} \text{g} ~.
\ee

\begin{itemize}
	
	\item $M_c$:
	
\end{itemize}

Then, let us see how long would a PBH with formation mass $M_f > M_q$ fall into $M_q \simeq 1.95 \times 10^{14}$ g:
\be
 t_q(M_f) \simeq \frac{M_f^3 - M_q^3}{\phi_*} 
 \equiv t_0 \frac{M_f^3 - M_q^3}{\bar{M}_*^3} ~.
\ee
If the formation mass $M_f$ is larger than a critical mass scale $M_c$, such a PBH would not reach $M_q$ at present, i.e., $t_q(M_f) > t_0$, which yields
\be
 M_c \equiv ( \bar{M}_*^3 + M_q^3 )^{1/3}
 \simeq 1.02 \bar{M}_*
 \simeq 5.17 \times 10^{14} \text{g} ~.
\ee
By using the range of $M_q$, and then some intermediate value at the last step. So that only PBHs slightly larger than $\bar{M}_*$ generate secondary emission at the present epoch. And the current mass is
\be
 M_0 = (M_f^3 - \bar{M}_*^3)^{1/3}, ~~~ \text{for} ~ M_f \geq M_c ~.
\ee

\begin{itemize}
 \item $M_*$:
\end{itemize}

PBHs completing their evaporation today have $M_0 =0$, the corresponding formation mass reads
\be
 M_* \simeq \Big[ \bar{M}_*^3 + ( 1 -\alpha^{-1} ) M_q^3 + ( \alpha^{-1} - \omega^{-1} ) M_g^3 \Big]^{1/3}
 \simeq 5.14 \times 10^{14} \text{g} ~.
\ee

\begin{figure}[h]
	\centering
	\includegraphics[width=3.5in]{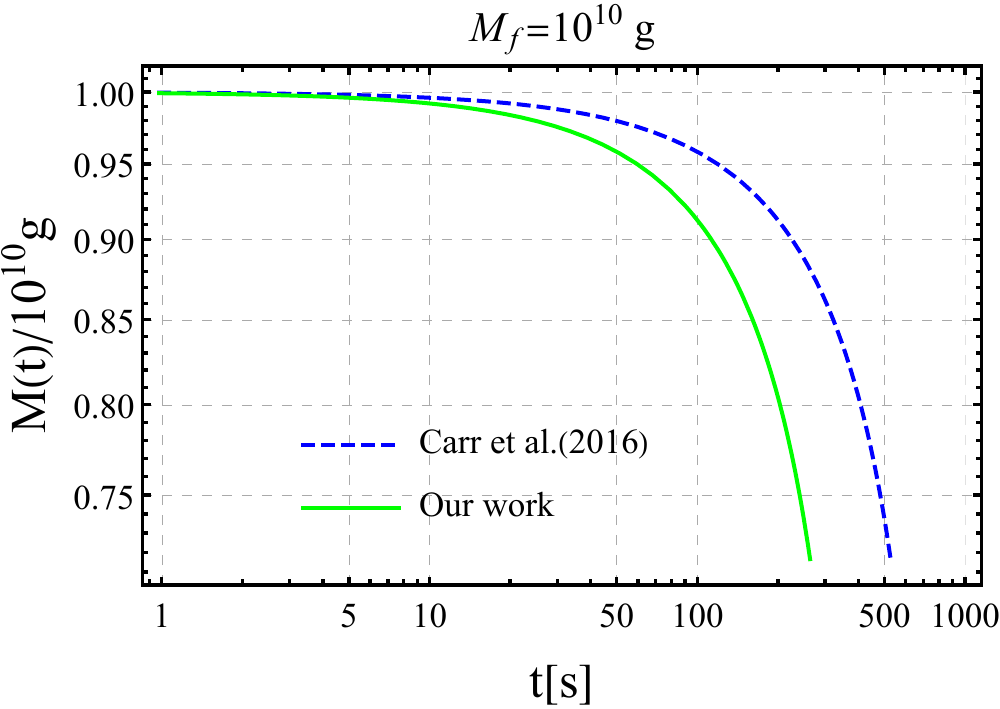}
	\includegraphics[width=3.5in]{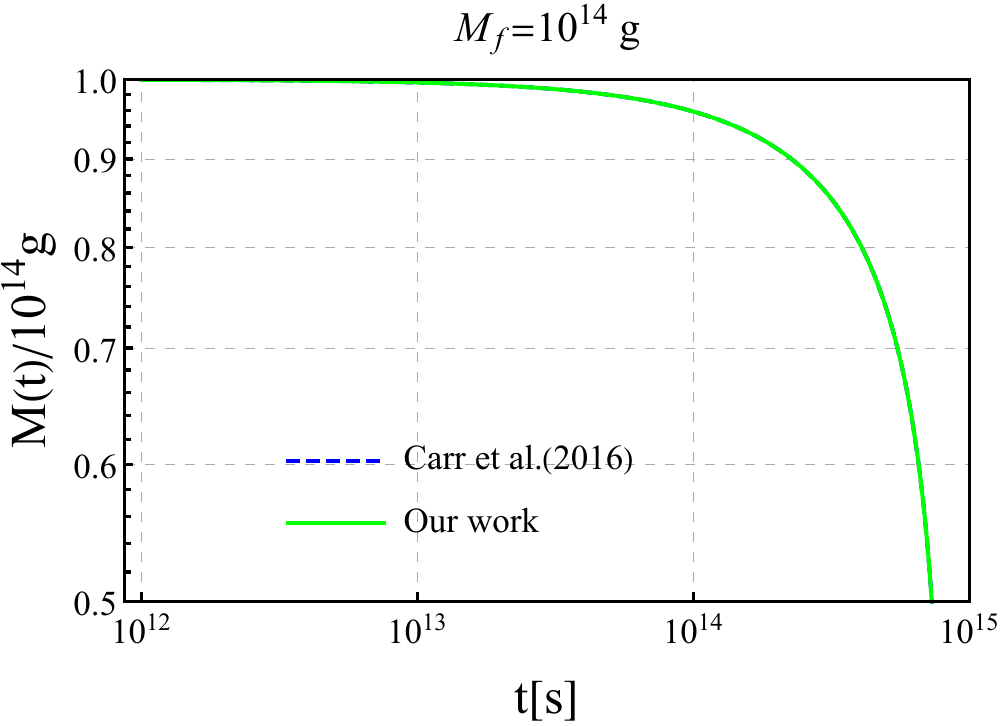}
	\caption{The comparisons between approximations used in Ref. \cite{Carr:2016hva} and Eq.  \eqref{phi_appro} for the mass evolutions of $M_f = 10^{10}, 10^{14}$ g, respectively. Our approximation \eqref{phi_appro} is more precise for the small formation mass $M_f$.}
	\label{fig:EvoMass}
\end{figure}

\begin{itemize}
	\item $M_1$:
\end{itemize}
The formation mass whose current mass is exact $M_g$:
\be
M_1 = \l[ \bar{M}_*^3 + ( 1 - \alpha^{-1} ) M_q^3 + \alpha^{-1} M_g^3 \r]^{1/3} ~.
\ee
We can see that the mass scales $M_1$ and $M_*$ are very close to each other, and it is expected that the new threshold mass $M_g$ should not affect the large-mass region significantly, which is confirmed by the following results, see Fig. \ref{fig:EvoMass}.

\begin{figure}[h]
	\centering
	\includegraphics[width=3.8in]{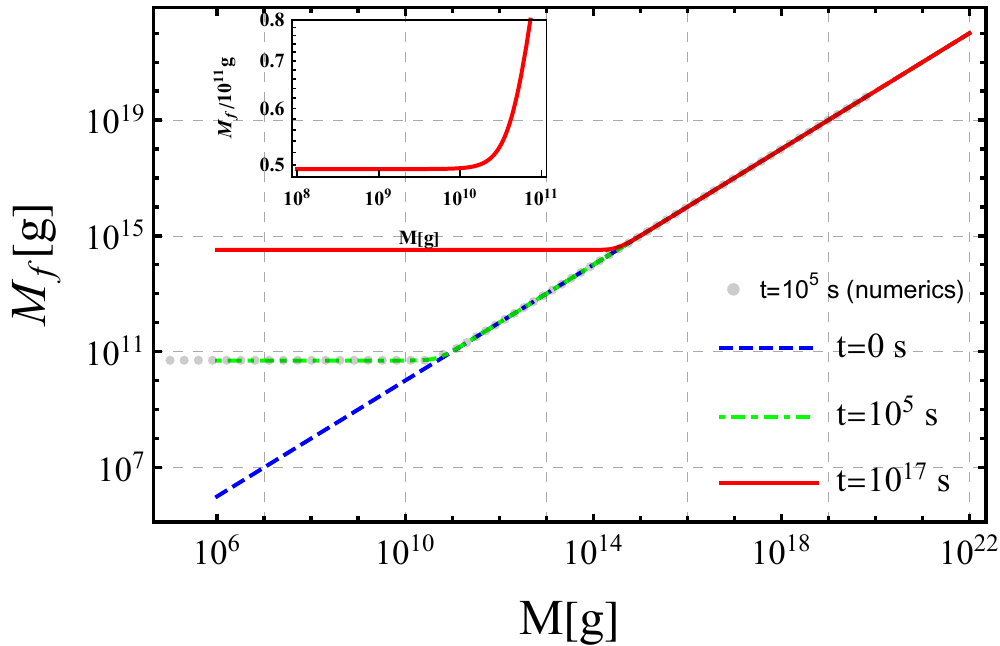}
	\caption{The relations between the formation mass $M_f$ and the mass $M$ at various cosmic times $t=\{ 0, 10^5, 10^{17} \}$ s after their formations, denoted by the blue dashed, green dashed and red solid curves, respectively. The grey dotted line is obtained numerically by the public code BlackHawk \cite{Arbey:2019mbc, Arbey:2021mbl}.}
	\label{fig:MfToM}
\end{figure}

With above preparations, we can calculate the expressions for a PBH's mass at time $t$ and its inverse $M(M_f, t)$ and $M_f(M, t)$, respectively, see Appendix \ref{app:mass}. The relation between $M_f$ and $M$ is shown in Fig. \ref{fig:MfToM}, where we set the cosmic times $t=\{ 0, 10^5, 10^{17} \}$ s after PBH formation. These curves are reasonable since those PBHs of $M_f \sim 10^{11}, 10^{15}$ g are at the end of their lifetimes around $t \sim 10^5, 10^{17}$ s, respectively.

The time evolution of mass function of PBHs is the interplay of the Hawking radiation and the cosmic expansion. So, it is convenient to start with the comoving mass function which is defined as
\be
 n_c(M(t)) \equiv {dn_c \over dM} ~,
\ee
where $dn_c$ is the comoving number density over the mass interval $(M, M+dM)$. Note that the time-evolving mass function is implicitly dependent on time through the evaporating mass $M(t)$. Using the chain rule for differentiation, we can relate the mass function at time $t$ to the formation mass function $n_c(M_f)$ in the following form:
\be
 n_c(M(t)) = \l( {dn_c \over dM_f} \r)_f {dM_f(M(t)) \over dM(t)} ~, 
\ee
and the explicit expressions are written as
\be \label{TimeEvolvingMassFunc}
n_c(M(t)) 
\simeq
\l( {dn_c \over dM_f} \r)_f(M) \times \lambda(M,t) ~,
\ee
where the transfer function is calculated as
\be \label{tranfer_func}
\begin{aligned}
	&\lambda(M,t) =
	\\&
	\l\{  	{\small
		\begin{aligned}
			& \l( 1 + \kappa {\bar{M}_*^3 \over M^3} {t \over t_0} \r)^{-2/3}, &&M \geq 10M_*
			\\
			& \l[ \kappa^{-1/2} + (\kappa^{-3/2} - \kappa^{-1/2}) { (10 M_*)^3 \over M^3 } + \kappa^{-1/2} { \bar{M}_*^3 \over M^3 } {t \over t_0} \r]^{-2/3}, && \l[ (10 M_*)^3 - \bar{M}_*^3 {t \over t_0} \r]^{1/3} \leq M \leq 10M_*
			\\
			& \l( 1 + {\bar{M}_*^3 \over M^3} {t \over t_0} \r)^{-2/3} , && M_q \leq M \leq \l[ (10 M_*)^3 - \bar{M}_*^3 {t \over t_0} \r]^{1/3}
			\\&
			\l[ \alpha^{1/2} + (\alpha^{3/2} - \alpha^{1/2}) { M_q^3 \over M^3} + \alpha^{3/2} { \bar{M}_*^3 \over M^3} {t \over t_0} \r]^{-2/3}, && \Theta\Big( t_0 {M_q^3 - M_g^3 \over \alpha \bar{M}_*^3 } - t \Big) \l( M_q^3 - \alpha \bar{M}_*^3 { t \over t_0} \r)^{1/3} 
			\\& \quad &&
			+ \Theta\Big( t - t_0 {M_q^3 - M_g^3 \over \alpha \bar{M}_*^3 } \Big) M_g \leq M \leq M_q
			\\
			& \l( 1 + \alpha {\bar{M}_*^3 \over M^3} {t \over t_0} \r)^{-2/3}, && M_g \leq M \leq \Theta\Big( t_0 {M_q^3 - M_g^3 \over \alpha \bar{M}_*^3 } - t \Big) \l( M_q^3 - \alpha \bar{M}_*^3 { t \over t_0} \r)^{1/3}
			\\
			& \Big[ \omega^{1/2} + (\omega^{3/2} - \omega^{3/2} \alpha^{-1}) {M_q^3 \over M^3} + (\omega^{3/2} \alpha^{-1} - \omega^{1/2} ) {M_g^3 \over M^3} && \Theta\Big(t - t_0 {M_q^3 - M_g^3 \over \alpha \bar{M}_*^3 } \Big) \Theta\Big(t_0 {\alpha^{-1} M_q^3 + (1 - \alpha^{-1}) M_g^3 \over \bar{M}_*^3 }  -t\Big)
			\\& \quad\quad\quad\quad \quad\quad\quad\quad \quad\quad\quad\quad
			+ {\bar{M}_*^3 \over M^3}{ t \over t_0 } \Big]^{-2/3}, &&  \l[ \omega \alpha^{-1} M_q^3 + (1 - \omega \alpha^{-1}) M_g^3 - \omega \bar{M}_*^3 { t \over t_0 } \r]^{1/3} \leq M \leq M_g
			\\&
			\Big[ \omega^{1/2} \alpha^{-1/2} + (\omega^{3/2} \alpha^{-3/2} - \omega^{1/2} \alpha^{-1/2}) {M_g^3 \over M^3} 
			&& \Theta\Big( t_0 {M_g^3 \over \omega \bar{M}_*^3 } - t \Big) \l( M_g^3 - \omega \bar{M}_*^3 { t \over t_0} \r)^{1/3} \leq M \leq
			\\& \quad\quad\quad\quad \quad\quad\quad\quad \quad\quad\quad\quad
			+ \omega^{3/2} \alpha^{-1/2} {\bar{M}_*^3 \over M^3} { t \over t_0 } \Big]^{-2/3}
			&& \Theta\Big(t - t_0 {M_q^3 - M_g^3 \over \alpha \bar{M}_*^3 } \Big) \Theta\Big(t_0 {\alpha^{-1} M_q^3 + (1 - \alpha^{-1}) M_g^3 \over \bar{M}_*^3 }  -t\Big) 
			\\& \quad
			&& \times \l[ \omega \alpha^{-1} M_q^3 + (1 - \omega \alpha^{-1}) M_g^3 - \omega \bar{M}_*^3 { t \over t_0 } \r]^{1/3} 
			\\& \quad
			&&+ \Theta\Big( t_0 {M_q^3 - M_g^3 \over \alpha \bar{M}_*^3 } - t \Big) M_g
			\\
			& \l( 1 + \omega {\bar{M}_*^3 \over M^3} {t \over t_0} \r)^{-2/3}, && M \leq \Theta\Big( t_0 {M_g^3 \over \omega \bar{M}_*^3 } - t \Big) \l( M_g^3 - \omega \bar{M}_*^3 { t \over t_0} \r)^{1/3}	
	\end{aligned} }
	\r.
	~.
\end{aligned}
\ee

Note that we do not need to take the differentiation to the Heaviside function $\Theta$, as which is introduced to take into account the different evaporation stages, see Eq. \eqref{phi_appro}. Then the physical mass function $n(M(t))$ is straightforward to calculate from the above,
\be
 n(M(t)) = a^3(t) n_c(M(t)) = n(M_f(M)) \times \lambda(M,t) ~,
\ee
where $n(M_f) = \l( {dn \over dM_f} \r)_f$ is the formation physical mass function and $\lambda(M,t)$ is shown in Eq.~\eqref{tranfer_func}.

\begin{figure}[h]
	\centering
	\includegraphics[width=3.8in]{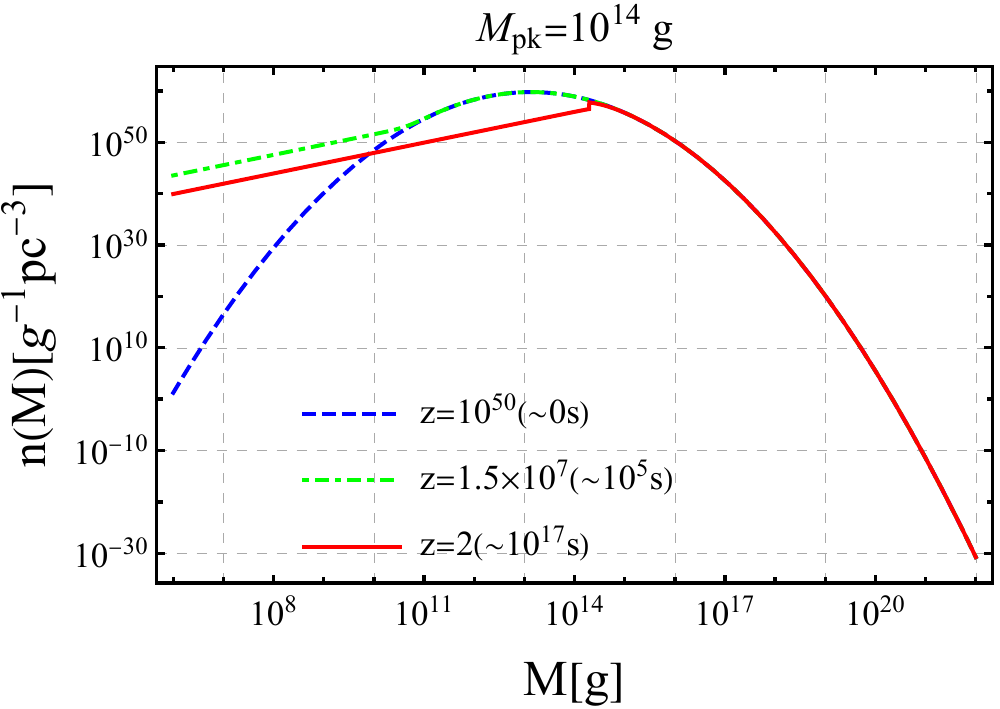}
	\caption{The mass functions at time $t = 0, 10^5, 10^{17}$ s for the initial lognormal mass functions, denoted by the blue dashed, green dashed and red solid curves, respectively. We choose the parameters for the initial mass function: $M_{pk} = 10^{14}$ g, $\beta = 10^{-23}$.}
	\label{fig:MassFunc}
\end{figure}

\begin{figure}[h]
	\centering
	\includegraphics[width=3.8in]{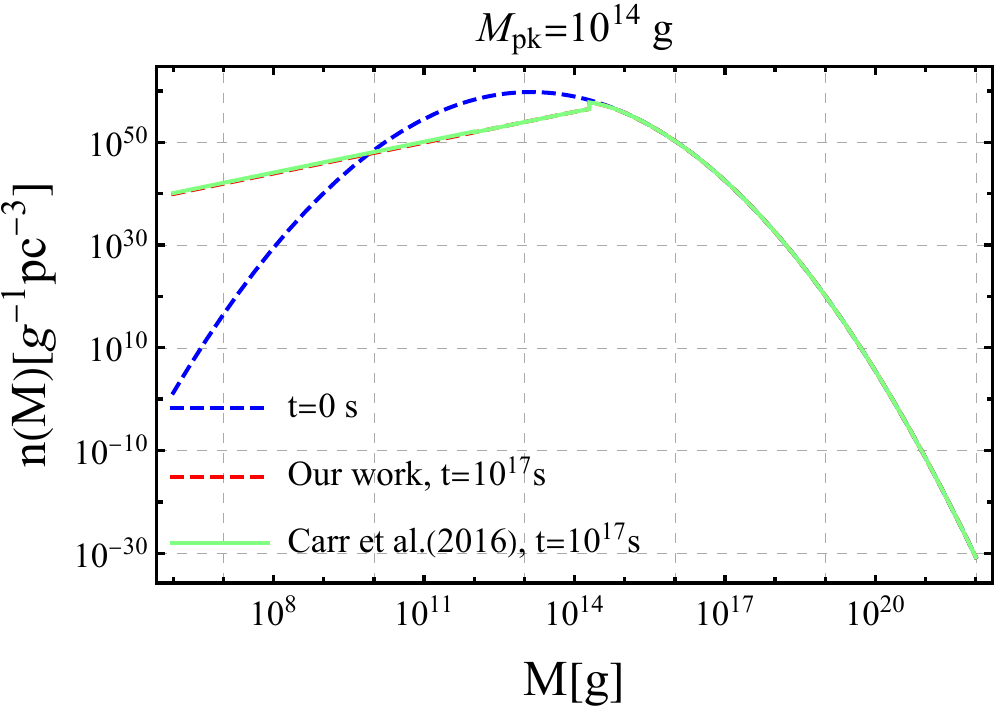}
	\caption{Two approximations used in Ref. \cite{Carr:2016hva} and Eq. \eqref{phi_appro} for the time-dependent mass function of $M_{pk} = 10^{14}$ g, respectively.}
	\label{fig:comparison}
\end{figure}

For the lognormal formation mass function,
\be \label{lognormal}
 n(M_f) = {\beta \rho_\text{tot} \over \sqrt{2 \pi} \sigma M_f^2} \exp\Big[  - { \ln^2(M_f/M_{pk}) \over 2 \sigma^2 }  \Big] ~,
\ee
which is most common mass distribution of PBHs and the other types of mass functions (i.e. the power-law and the critical ones) can be rewritten in the form of lognormal distribution \cite{Carr:2017jsz}. The quantity $\beta \equiv \rho_\text{PBH} / \rho_\text{tot}$ is the formation abundance of PBHs, and the total energy density of Universe at the formation epoch that can be calculated by the Friedmann equation, i.e., $\rho_\text{tot}(z_f) = 3 M_\text{pl}^2 H^2(z_f)$, $H^2(z) = H_0^2 [ \Omega_r (1+z)^{4} + \Omega_k (1+z)^{2} + \Omega_m (1+z)^{3} + \Omega_\Lambda ]$, here $\Omega_r \simeq 10^{-4}$, $\Omega_k \simeq 0$, $\Omega_m \simeq 0.315$ and $\Omega_\Lambda \simeq 0.6847$ are normalized radiation, curvature, baryon and dark energy density parameters, respectively. $z_f$ is the formation redshift. For the purpose of numerical computation, we normalize the present scale factor $a(t_0) = 1$, and chose the Hubble parameter $H_0 = 67.4\ \text{km}~ \text{s}^{-1}~ \text{Mpc}^{-1}$. Using the redshift-time relation $dt = - { dz \over H(z) (1+z) }$, we can transfer the argument of any time-evolving function from the cosmic time $t$ to the redshift $z$. By choosing parameters $\beta = 10^{-23}$, $\sigma = 1$, $M_{pk} =10^{14}$ g, we plot the mass functions at different times $t = 0, 10^5, 10^{17}$ s originated from the initial lognormal mass function, which is shown in Fig. \ref{fig:MassFunc}. Fig.~\ref{fig:comparison} plots the comparison between approximations used in \cite{Carr:2016hva} and Eq. \eqref{phi_appro} for the lognormal mass distribution, one can easily see that they are quite close to each other.

\section{Test of Cosmological Models via Standard Timer} \label{sec:TestLCDM}

The cosmological redshift depends on the evolution of the scale factor $a(t)$, and the cosmological time is the time elapsed from the primordial physical scale to the present physical scale which follows $t \sim H_0^{-1}$. Hence, a cosmological model leaves the imprint on the redshift-time calibration which indicates that the cosmological parameters can be constrained by the standard timer.

From the definition of cosmological redshift $1 + z = a_0/a$ and the Hubble parameter $H(z) = \dot{a}/a$, we obtain the redshift-time relation,
\begin{align}\label{eq:redtime}
	\frac{dz}{(1+z) H(z)} = - dt~.
\end{align}
Then, the redshift-time calibration $(z_0, z_\mathrm{PBH}, \Delta t)$ between the PBH bubble and the redshift calibrator puts constraint on the Hubble parameter evolution as following,
\begin{align}
	\int_{z_0}^{z_\mathrm{PBH}} \frac{dz}{(1+z) H(z)} = \int_{t_\mathrm{PBH}}^{t_0} dt = \Delta t~.
\end{align}
Considering the flat $\Lambda \mathrm{CDM}$ model as an example, $H(z)$ can be expressed in following form,
\begin{align}\label{eq:hubblez}
	H(z) = H_0 \sqrt{\Omega_\gamma (1 + z)^4 + \Omega_m (1 + z)^3 + \Omega_\Lambda}~.
\end{align}
After extracting redshift-time calibration from a bunch of PBH bubbles covering from low redshifts to high redshifts, we can apply the Markov chain Monte Carlo (MCMC) simulation on flat $\Lambda \mathrm{CDM}$ model to constrains the cosmological parameters $H_0$, $\Omega_\gamma$, $\Omega_m$ and $\Omega_\Lambda$, 

In particular, we can evaluate the average Hubble parameter for two nearby redshifts $z_1$ and $z_2$, where $z_1 \sim z_2$. In Eq.~\eqref{eq:redtime}, we write
\begin{align}
	\ln(\frac{1+z_2}{1+z_1}) = \ln(\eta) = \bar{H}(\frac{z_1 + z_2}{2}) \Delta t~.
\end{align}
Then we yield
\begin{align}
	\bar{H}(\frac{z_1 + z_2}{2}) = \frac{\ln \eta}{\Delta t}~.
\end{align}

\section{Conclusion and Discussions}\label{sec:concl}

To conclude, we present a new approach to track the historical evolution of the Universe, which is named as the standard timer. We illustrate how the standard timer works by constructing the calibration between the redshift and physical time from the observed EM signals of PBH bubbles. The feasibility that PBH bubble can help track the evolution of the Universe is that their identical primordial origin produces the same initial mass function inside the bubble at the same epoch, then the redshift and the physical time calibration can be decoded from the observed EM signals and the internal mass function evolution. Also, the existence of PBH bubbles in the primordial Universe extends the narrow redshift windows in standard candle, standard ruler and standard siren, which makes the valid redshift range of standard timer cover from the primordial epoch to the present. As we have shown in Fig.~1 of \cite{Cai:2021zxo}, the EM signals from a PBH bubble with mass $M_\mathrm{bub} = 10^{38} \mathrm{g}$ can be detected up $z \sim 10^3$.

In developing the method of standard timers, the key step is to calibrate the redshift and physical evolution time for two PBH bubbles. Following \cite{1142844, PhysRevLett.64.1193}, we apply the inverse problem to Hawking radiation, by the least square method in Eq.~\eqref{eq:leastsquare}, we can extract the redshifted mass function from observed EM signal spectrum. Then redshift ratio $\eta$ can be obtained by matching redshifted mass functions for two PBH bubbles with error around $0.03 \%$. Applying the obtained redshift ratio into evolution of mass function in Eq.~\eqref{eq:lowmass}, the calibration between redshift and physical time can be constructed with error around $0.2 \%$ in Fig.~\ref{fig:red_error} and $0.1 \%$ in Fig.~\ref{fig:time_error}, respectively. This ensures the high precision of the standard timer.

Practically, various uncertainties should be taken into account in the observed EM signals from PBH bubbles, e.g., systematic uncertainties from the limited optical depth of the ultra-high gamma ray \cite{nikishov1961absorption}, measurement uncertainties in the gamma-ray spectrometry \cite{L_py_2015}, etc. In data analysis, some regularization methods \cite{PROVENCHER1982213} in inverse problems should be applied to increase the accuracy of standard timer. In particular, under the low mass approximation, $n(M) \sim M^2$ in evolution of PBH mass function, which produces $F(E) \sim E^{-1}$ in the high-energy spectrum. This provides a specific template for filtering out the noise in gamma ray signals and evidence for potential PBH bubble candidate in gamma ray sources.  

In general, there could exist other types of primordial phenomena that can track the evolution of the Universe, such as collision of cosmic strings \cite{Shellard:1987bv} and domain walls \cite{Takamizu:2004rq}. The connection between their primordial origin and the observed signals could build the one-to-one mapping between the internal physical evolution and the external evolution of the Universe, which are the potential tools in studying the history of the Universe.

\section*{Acknowledgement}

This work is supported in part by the National Key R\&D Program of China (2021YFC2203100). YFC is supported in part by the NSFC (Nos. 11961131007, 11653002), by the National Youth Talents Program of China, by the Fundamental Research Funds for Central Universities, by the CSC Innovation Talent Funds, by the CAS project for Young Scientists in Basic Research (YSBR-006), and by the USTC Fellowship for International Cooperation. CC, QD and YW are supported in part by the CRF grant C6017-20GF, the GRF grant 16303621 by the RGC of Hong Kong SAR, and the NSFC Excellent Young Scientist Scheme (Hong Kong and Macau) Grant No.~12022516.
We acknowledge the use of computing facilities of HKUST, as well as the clusters {\it LINDA} and {\it JUDY} of the particle cosmology group at USTC.

\appendix

\section{Inverse Black-body Radiation Problem} \label{inverseBlackbody}

Given a radiation spectrum $P(v)$ from the black-body radiator, to find the temperature distribution $a(T)$ on the blackbody, i.e., the so called inverse black-body radiation problem. The radiation spectrum from the black-body radiator can be expressed as
\begin{align}\label{eq:spectrum}
P(v) = \frac{2 h v^3}{c^2}\int_0^{\infty} \frac{a(T)}{e^{hv/kT} - 1} dT~.
\end{align}
Here, $h$ is Planck constant, $k$ is the Boltzmann constant, $c$ is the speed of light. In following discussion, we use two methods to resolve the temperature distribution $a(T)$.

\subsection{Iteration method}
We follow the iteration method investigated in Ref. \cite{1142844}. First, we introduce the coldness which is defined as $u \equiv h/kT$, then, Eq.~\eqref{eq:spectrum} can be rewritten as
\begin{align}\label{eq:spectrum_v2}\nonumber
	P(v) &= \frac{2 h v^3}{c^2}\int_0^{\infty} \frac{a(h/ku)}{e^{uv} - 1} \frac{h}{k} \frac{1}{u^2} du\\
		 &= \frac{2 h v^3}{c^2}\int_0^{\infty} \frac{b(u)}{e^{uv} - 1} du~.
\end{align}
Here, we define $b(u) \equiv \frac{h}{k u^2} a(\frac{h}{ku})$.Using the series expansion:
\begin{align}\label{eq.series}
	\frac{1}{e^{uv} - 1} = e^{-uv}\frac{1}{1-e^{-uv}} = e^{-uv}\sum_{n = 0}^{\infty}e^{-nuv} = \sum_{n = 1}^{\infty}e^{-nuv}~.
\end{align}
we yield
\begin{align}\label{eq:spectrum_v3}\nonumber
	P(v) &= \frac{2 h v^3}{c^2}\int_0^{\infty} \sum_{n = 1}^{\infty} e^{-nuv} b(u) du\\\nonumber
		 &= \frac{2 h v^3}{c^2}\int_0^{\infty} e^{-\tilde{u} v} \sum_{n = 1}^{\infty} \frac{1}{n} b(\frac{\tilde{u}}{n}) d\tilde{u}\\\nonumber
		 &= \frac{2 h v^3}{c^2}\int_0^{\infty} e^{-uv} f(u) du\\
		 &= \frac{2 h v^3}{c^2}\mathcal{L}(f(u)), ~~~~~~~\tilde{u} \equiv n u, ~~f(u)\equiv \sum_{n = 1}^{\infty} \frac{1}{n} b(\frac{u}{n}) ~.
\end{align}
Here $\mathcal{L}(f)$ is the Laplace transformation of $f$. The function $f(u)$ is given by
\begin{align}
	f(u) = \mathcal{L}^{-1}(\frac{c^2}{2 h v^3}P(v))~.
\end{align}
In order to obtain $b(u)$, the iteration method can be applied to $f(u)$, which is expressed as
\begin{align}
	f(u) = \sum_{n = 1}^{\infty} \frac{1}{n} b(\frac{u}{n}) = b(u) + \sum_{n = 2}^{\infty} \frac{1}{n} b(\frac{u}{n})~,\\
	b_{m+1}(u) = f(u) - \sum_{n = 2}^{\infty} \frac{1}{n} b_{m}(\frac{u}{n})~.
\end{align}
Then $b(u)$ can be obtained by
\begin{align}
	b(u) = \lim_{m \rightarrow \infty}b_{m}(u)~.
\end{align}
The temperature distribution $a(T)$ is
\begin{align}
	a(T) = \frac{h}{k T^2} b(\frac{h}{kT})~.
\end{align}

\subsection{M\"{o}bius inversion transformation}

We follow Ref. \cite{PhysRevLett.64.1193} for M\"{o}bius inversion transformation method. Given a function:
\begin{align}
	g(x) = \sum_{n = 1}^{\infty} f(n x)~,
\end{align}
where $n$ are integers. We derive the M\"{o}bius inversion formula as follows
\begin{align}\label{eq:mobius}
	f(x) = \sum_{n = 1}^{\infty} \mu(n) g(n x)~.
\end{align}
Here, $\mu(n)$ is the M\"{o}bius function defined as
\begin{align}\label{dEnergy}
	& \mu(n) =
	\left\{
	\begin{aligned}
		&1 \qquad  ~,~n = 1 \\
		&(-1)^s ~,~n = p_1 p_2 p_3...p_s \\
		&0 \qquad ~,~\mathrm{otherwise}
	\end{aligned}
	\right. .
\end{align}
Here, $p_s$ is $s$th-order prime factor. Then, we can apply the M\"{o}bius inversion formula on inverse black-body radiation problem as follows,
\begin{align}\label{eq:laplace}\nonumber
	P(v) &= \frac{2 h v^3}{c^2}\int_0^{\infty} \sum_{n = 1}^{\infty} e^{-nuv} b(u) du\\\nonumber
		 &= \frac{2 h v^3}{c^2}\sum_{n = 1}^{\infty} \int_0^{\infty} e^{-nuv} b(u) du\\
		 &= \frac{2 h v^3}{c^2}\sum_{n = 1}^{\infty} \mathcal{L}(b(u); u \rightarrow nv)~.
\end{align}
Combining Eqs.~\eqref{eq:mobius} and \eqref{eq:laplace}, we obtain the Laplace transformation of function $b(u)$:
\begin{align}
	\mathcal{L}(b(u); u \rightarrow v) = \sum_{n = 1}^{\infty} \mu(n) \frac{c^2}{2 h} \frac{P(n v)}{(n v)^3}~,
\end{align}
which gives
\begin{align}
	b(u) = \frac{c^2}{2 h} \sum_{n = 1}^{\infty} \frac{\mu(n)}{n^3} \mathcal{L}^{-1}(\frac{P(n v)}{v^3}; v \rightarrow u)~.
\end{align}
The temperature distribution $a(T)$ is
\begin{align}
	a(T) = \frac{h}{k T^2} b(\frac{h}{kT})~.
\end{align}
\section{Formal Method for Inverse Hawking Radiation Problem}\label{inverseHawking}

A black hole can emit particles similar to the black-body radiation with energies in the range $(E, E + dE)$ at a rate \cite{Hawking:1974rv, Hawking:1974sw}
\begin{align}
	\frac{d^2N}{dt dE} = \frac{1}{2 \pi} \frac{\Gamma_s(E,M)}{e^{8 \pi G M E} - (-1)^{2s}} n_\text{dof} ~,
\end{align}
where $n_\text{dof} = 2$ for photon $(s = 1)$ and $\Gamma_s(E, M)$ can be expressed as \cite{MacGibbon:1990zk}
\begin{align} \label{greyfactor}
&\Gamma_1(E,M) \propto
\left\{
\begin{aligned}
& G^4 M^4 E^4 ~, ~E < (8 \pi G M)^{-1} , \\
& G^2 M^2 E^2 ~, ~E > (8 \pi G M)^{-1} .
\end{aligned}
\right. 
\end{align}
For a given PBH mass function $n(M)$, we can obtain photon number density emission rate as follows:
\begin{align}\label{eq:hawking_spectrum}
	\frac{d^2 n }{dt dE} = \int_{0}^{\infty} \frac{d^2 N}{dt dE} n(M) dM~.
\end{align}
Similar to the inverse black-body radiation problem, given an emission rate spectrum, finding the PBH mass function $n(M)$, is called the inverse Hawking radiation problem. The crucial difference between Hawking radiation and black-body radiation is the $\Gamma_s(E, M)$ term, here we denote particle emission rate $R(E) \equiv d^2 n / dt dE$ and consider the photon with $s = 1$, the \eqref{eq:hawking_spectrum} can be written as
\begin{align}\nonumber
R(E)&= \frac{1}{2 \pi} \int_{0}^{\infty}  \frac{\Gamma_1(E,M)}{e^{8 \pi G M E} - 1} n(M) dM\\\nonumber
&= \frac{ A G^4 }{2 \pi} \int_{0}^{\frac{1}{8 \pi G E}}  \frac{E^4 M^4}{e^{8 \pi G M E} - 1} n(M) dM 
+ \frac{B G^2}{2 \pi} \int_{\frac{1}{8 \pi G E}}^{\infty}  \frac{E^2 M^2}{e^{8 \pi G M E} - 1} n(M) dM\\\nn 
&~~~~+ \frac{B G^2}{2 \pi} \int_{0}^{\frac{1}{8 \pi G E}}  \frac{E^2 M^2}{e^{8 \pi G M E} - 1} n(M) dM
- \frac{B G^2}{2 \pi} \int_{0}^{\frac{1}{8 \pi G E}}  \frac{E^2 M^2}{e^{8 \pi G M E} - 1} n(M) dM\\
&= \frac{ A G^4 }{2 \pi} \int_{0}^{\frac{1}{8 \pi G E}}  \frac{E^2 M^2 (E^2 M^2 - {B \over A  G^2})}{e^{8 \pi G M E} - 1} n(M) dM 
+ \frac{B G^2}{2 \pi} \int_{0}^{\infty}  \frac{E^2 M^2}{e^{8 \pi G M E} - 1} n(M) dM ,
\end{align}
where $A$ and $B$ are coefficients in the grey factor \eqref{greyfactor}. Therefore, we yield the relation:
\begin{align}\label{eq:hawking_inverse}
\frac{B G^2}{2 \pi} \int_{0}^{\infty} \frac{E^2 M^2}{e^{8 \pi G M E} - 1} n(M) dM 
= R(E) + \frac{ A G^4 }{2 \pi} \int_{0}^{\frac{1}{8 \pi G E}}  \frac{E^2 M^2 ({B \over A  G^2} - E^2 M^2 )}{e^{8 \pi G M E} - 1} n(M) dM ~.
\end{align}
In order to obtain the mass function $n(M)$, we use the iteration method:
\begin{align}\label{eq:hawking_iteration}
\frac{B G^2}{2 \pi} \int_{0}^{\infty}  \frac{E^2 M^2}{e^{8 \pi G M E} - 1} n_{m+1}(M) dM 
= R(E) + \frac{ A G^4 }{2 \pi} \int_{0}^{\frac{1}{8 \pi G E}}  \frac{E^2 M^2 ({B \over A  G^2} - E^2 M^2 )}{e^{8 \pi G M E} - 1} n_{m}(M) dM ~.
\end{align}
Obviously , the iteration method works for the condition:
\begin{align} 
\lim_{m \rightarrow \infty}n_{m}(M) = n(M)~.
\end{align}
So we need to check the convergence of the iteration relation Eq.~\eqref{eq:hawking_iteration}, we replace $n_{m}(M)$ with $n(M) + \epsilon_{m}(M)$ and yield
\begin{align}\label{eq:hawking_error_iteration}\nonumber
&~~~~\frac{B G^2}{2 \pi} \int_{0}^{\infty}  \frac{E^2 M^2}{e^{8 \pi G M E} - 1} n(M) dM
+ \frac{B G^2}{2 \pi} \int_{0}^{\infty}  \frac{E^2 M^2}{e^{8 \pi G M E} - 1} \epsilon_{m+1}(M) dM \\
&= R(E) + \frac{ A G^4 }{2 \pi} \int_{0}^{\frac{1}{8 \pi G E}}  \frac{E^2 M^2 ({B \over A  G^2} - E^2 M^2 )}{e^{8 \pi G M E} - 1} n(M) dM 
+ \frac{ A G^4 }{2 \pi} \int_{0}^{\frac{1}{8 \pi G E}}  \frac{E^2 M^2 ({B \over A  G^2} - E^2 M^2 )}{e^{8 \pi G M E} - 1} \epsilon_{m}(M) dM~.
\end{align}
Applying Eq. \eqref{eq:hawking_inverse} into Eq. \eqref{eq:hawking_error_iteration}, we obtain
\begin{align}\label{eq:hawking_unequal}\nonumber
\frac{B G^2}{2 \pi} \int_{0}^{\infty}  \frac{E^2 M^2}{e^{8 \pi G M E} - 1} \epsilon_{m+1}(M) dM &= \frac{ A G^4 }{2 \pi} \int_{0}^{\frac{1}{8 \pi G E}}  \frac{E^2 M^2 ({B \over A  G^2} - E^2 M^2 )}{e^{8 \pi G M E} - 1} \epsilon_{m}(M) dM\\\nonumber
&< \frac{B G^2}{2 \pi} \int_{0}^{\frac{1}{8 \pi G E}}  \frac{E^2 M^2 }{e^{8 \pi G M E} - 1} \epsilon_{m}(M) dM\\
&< \omega \frac{B G^2}{2 \pi} \int_{0}^{\infty}  \frac{E^2 M^2 }{e^{8 \pi G M E} - 1} \epsilon_{m}(M) dM .
\end{align}
Here, it is easy to find a positive real number $\omega < 1$ to satisfy the inequality of Eq. \eqref{eq:hawking_unequal}, we therefore prove that
\begin{align} 
\lim_{m \rightarrow \infty}\epsilon_{m}(M) = 0~.
\end{align}
and
\begin{align} 
\lim_{m \rightarrow \infty} \frac{1}{2 \pi} \int_{0}^{\infty}  \frac{E^2 M^2}{e^{8 \pi G M E} - 1} n_{m+1}(M) dM = \frac{1}{2 \pi} \int_{0}^{\infty}  \frac{E^2 M^2}{e^{8 \pi G M E} - 1} n(M) dM ~.
\end{align}
Hence, the mass function $n(M)$ can be solved by iteration method and M\"obius inversion transformation we have discussed above.

\section{The Analytic Formulas for $M(t)$ and $M_f(M)$}\label{app:mass}

As discussed in Sec. \ref{sec:MassFunc}, for the large formation mass $M_f$, the time evolution remains unchanged, i.e., 
\be
M(t) \simeq \l( M_f^3 - \kappa \bar{M}_*^3 {t \over t_0} \r)^{1/3},~ M_f \geq M_a~ ,
\ee
\bl
M(t) 
\simeq&
\Theta\Big( t_0 \frac{M_f^3 - (10 M_*)^3 }{ \kappa \bar{M}_*^3} - t \Big) \l( M_f^3 - \kappa \bar{M}_*^3 {t \over t_0} \r)^{1/3} \nn
\\&
+ \Theta\Big( t - t_0 \frac{M_f^3 - (10 M_*)^3}{ \kappa \bar{M}_*^3} \Big)
\l[ \kappa^{-1} M_f^3 + (1 - \kappa^{-1} ) (10 M_*)^3 - \bar{M}_*^3 {t \over t_0} \r]^{1/3},~ 10 M_* \leq M_f \leq M_a ~,
\el
\be
M(t) \simeq \l( M_f^3 - \bar{M}_*^3 {t \over t_0} \r)^{1/3},~ M_c \leq M_f \leq 10 M_* ~.
\ee
\bl
M(t) 
\simeq&
\Theta\Big( t_0 {M_f^3 - M_q^3 \over \bar{M}_*^3} - t \Big) \l( M_f^3 - \bar{M}_*^3 {t \over t_0} \r)^{1/3} \nn
\\&
+ \Theta\Big( t - t_0 {M_f^3 - M_q^3 \over \bar{M}_*^3} \Big)
\l[ \alpha M_f^3 + (1 - \alpha ) M_q^3 - \alpha \bar{M}_*^3 {t \over t_0} \r]^{1/3},~ M_1 \leq M_f \leq M_c ~,
\el
\bl
M(t) 
\simeq&
\Theta\Big( t_0 {M_f^3 - M_q^3 \over \bar{M}_*^3} - t \Big) \l( M_f^3 - \bar{M}_*^3 {t \over t_0} \r)^{1/3} \nn
\\&
+ \Theta\Big( t - t_0 {M_f^3 - M_q^3 \over \bar{M}_*^3} \Big) \Theta\Big( t_0 {M_f^3 + (\alpha^{-1} - 1) M_q^3 -  \alpha^{-1} M_g^3 \over \bar{M}_*^3} - t \Big)
\\& \quad\quad\quad\quad \quad\quad\quad\quad  \quad\quad\quad\quad
\times \l[ \alpha M_f^3 + (1 - \alpha ) M_q^3 - \alpha \bar{M}_*^3 {t \over t_0} \r]^{1/3} \nn
\\&
+ \Theta\Big( t - t_0 {M_f^3 + (\alpha^{-1} - 1) M_q^3 -  \alpha^{-1} M_g^3 \over \bar{M}_*^3} \Big) \nn
\\& \quad\quad\quad\quad \quad\quad\quad\quad  \quad\quad\quad\quad
\times \l[ \omega M_f^3 + \omega (\alpha^{-1} - 1 ) M_q^3 + (1 - \omega \alpha^{-1}) M_g^3 - \omega \bar{M}_*^3 {t \over t_0} \r]^{1/3}, ~ \tilde{M}_* \leq M_f \leq M_1
~,
\el
\bl
M(t) 
\simeq&
\Theta\Big( t_0 {M_f^3 - M_q^3 \over \bar{M}_*^3} - t \Big) \l( M_f^3 - \bar{M}_*^3 {t \over t_0} \r)^{1/3} \nn
\\&
+ \Theta\Big( t - t_0 {M_f^3 - M_q^3 \over \bar{M}_*^3} \Big) \Theta\Big( t_0 {M_f^3 + (\alpha^{-1} - 1) M_q^3 -  \alpha^{-1} M_g^3 \over \bar{M}_*^3} - t \Big)
\\& \quad\quad\quad\quad \quad\quad\quad\quad  \quad\quad\quad\quad
\times \l[ \alpha M_f^3 + (1 - \alpha ) M_q^3 - \alpha \bar{M}_*^3 {t \over t_0} \r]^{1/3} \nn
\\&
+ \Theta\Big( t - t_0 {M_f^3 + (\alpha^{-1} - 1) M_q^3 -  \alpha^{-1} M_g^3 \over \bar{M}_*^3} \Big) \nn
\\& \quad\quad\quad\quad \quad\quad\quad\quad
\times \Theta\Big( t_0 {M_f^3 + (\alpha^{-1} - 1) M_q^3 + ( \omega^{-1} - \alpha^{-1} ) M_g^3 \over \bar{M}_*^3} - t \Big)
\nn
\\& \quad\quad\quad\quad \quad\quad\quad\quad  \quad\quad\quad\quad
\times \l[ \omega M_f^3 + \omega (\alpha^{-1} - 1 ) M_q^3 + (1 - \omega \alpha^{-1}) M_g^3 - \omega \bar{M}_*^3 {t \over t_0} \r]^{1/3},~ M_q \leq M_f \leq \tilde{M}_*
~,
\el
\bl
M(t) 
\simeq&
\Theta\Big( t_0 {M_f^3 - M_g^3 \over \alpha \bar{M}_*^3} - t \Big) \l( M_f^3 - \alpha \bar{M}_*^3 {t \over t_0} \r)^{1/3} \nn
\\&
+ \Theta\Big( t - t_0 {M_f^3 - M_g^3 \over \alpha \bar{M}_*^3} \Big) \Theta\Big( t_0 { \alpha^{-1} M_q^3 + ( \omega^{-1} - \alpha^{-1}) M_g^3 \over \alpha \bar{M}_*^3} - t \Big) \nn
\\& \quad\quad\quad\quad \quad\quad\quad\quad  \quad\quad\quad\quad
\times
\l[ \omega \alpha^{-1} M_f^3 + (1 - \omega \alpha^{-1}) M_g^3 - \omega \bar{M}_*^3 {t \over t_0} \r]^{1/3},~ M_g \leq M_f \leq M_q
~,
\el
\be
M(t)
\simeq
\Theta\Big( t_0 { M_f^3 \over \omega \bar{M}_*^3 }- t \Big) \l( M_f^3 - \omega \bar{M}_*^3 {t \over t_0} \r)^{1/3},~ M_f \leq M_g ~.
\ee

The next step is to express $M_f$ in terms of $M$ and $t$.
\be
M_f \simeq \l( M^3 + \kappa \bar{M}_*^3 {t \over t_0} \r)^{1/3},~ M \geq 10M_* ~,
\ee

\be
M_f \simeq \l[ \kappa M^3 + (1 - \kappa) (10 M_*)^3 + \kappa \bar{M}_*^3 {t \over t_0} \r]^{1/3},~ \l[ (10 M_*)^3 - \bar{M}_*^3 {t \over t_0} \r]^{1/3} \leq M \leq 10M_* ~,
\ee

\be
M_f \simeq \l( M^3 + \bar{M}_*^3 {t \over t_0} \r)^{1/3} ,~ M_q \leq M \leq \l[ (10 M_*)^3 - \bar{M}_*^3 {t \over t_0} \r]^{1/3} ~,
\ee

\bl\nonumber
M_f &\simeq \l[ \alpha^{-1} M^3 + (1 - \alpha^{-1}) M_q^3 + \bar{M}_*^3 { t \over t_0 } \r]^{1/3}~,\\
\Theta & \Big( t_0 {M_q^3 - M_g^3 \over \alpha \bar{M}_*^3 } - t \Big) \l( M_q^3 - \alpha \bar{M}_*^3 { t \over t_0} \r)^{1/3} + \Theta\Big( t - t_0 {M_q^3 - M_g^3 \over \alpha \bar{M}_*^3 } \Big) M_g \leq M \leq M_q ~,
\el

\be
M_f \simeq \l( M^3 + \alpha \bar{M}_*^3 {t \over t_0} \r)^{1/3},~ M_g \leq M \leq \Theta\Big( t_0 {M_q^3 - M_g^3 \over \alpha \bar{M}_*^3 } - t \Big) \l( M_q^3 - \alpha \bar{M}_*^3 { t \over t_0} \r)^{1/3} ~.
\ee

\bl
M_f & \simeq \l[ \omega^{-1} M^3 + (1 - \alpha^{-1} ) M_q^3 + (\alpha^{-1} - \omega^{-1}) M_g^3 + \bar{M}_*^3 { t \over t_0 } \r]^{1/3},~ \nn
\\ \Theta \Big(t - t_0  {M_q^3 - M_g^3 \over \alpha \bar{M}_*^3 } \Big) & \Theta\Big(t_0 {\alpha^{-1} M_q^3 + (1 - \alpha^{-1}) M_g^3 \over \bar{M}_*^3 }  -t\Big) \l[ \omega \alpha^{-1} M_q^3 + (1 - \omega \alpha^{-1}) M_g^3 - \omega \bar{M}_*^3 { t \over t_0 } \r]^{1/3} \leq M \leq M_g ~,
\el
for the condition 
\bl\nonumber
&\Theta\Big( t_0 {M_g^3 \over \omega \bar{M}_*^3 } - t \Big) \l( M_g^3 - \omega \bar{M}_*^3 { t \over t_0} \r)^{1/3} \leq M
\leq \Theta\Big(t - t_0 {M_q^3 - M_g^3 \over \alpha \bar{M}_*^3 } \Big) \Theta\Big(t_0 {\alpha^{-1} M_q^3 + (1 - \alpha^{-1}) M_g^3 \over \bar{M}_*^3 }  -t\Big) \\
&\times \l[ \omega \alpha^{-1} M_q^3 + (1 - \omega \alpha^{-1}) M_g^3 - \omega \bar{M}_*^3 { t \over t_0 } \r]^{1/3}+ \Theta\Big( t_0 {M_q^3 - M_g^3 \over \alpha \bar{M}_*^3 } - t \Big) M_g,
\el
\be
M_f \simeq \l[ \omega^{-1} \alpha M^3 + (1 - \omega^{-1} \alpha) M_g^3 + \alpha \bar{M}_*^3 { t \over t_0 } \r]^{1/3} ~,
\ee

\be
M_f \simeq \l( M^3 + \omega \bar{M}_*^3 {t \over t_0} \r)^{1/3},~ M \leq \Theta\Big( t_0 {M_g^3 \over \omega \bar{M}_*^3 } - t \Big) \l( M_g^3 - \omega \bar{M}_*^3 { t \over t_0} \r)^{1/3} ~.
\ee

\begin{figure}[h]
	\centering
	\includegraphics[width=3.8in]{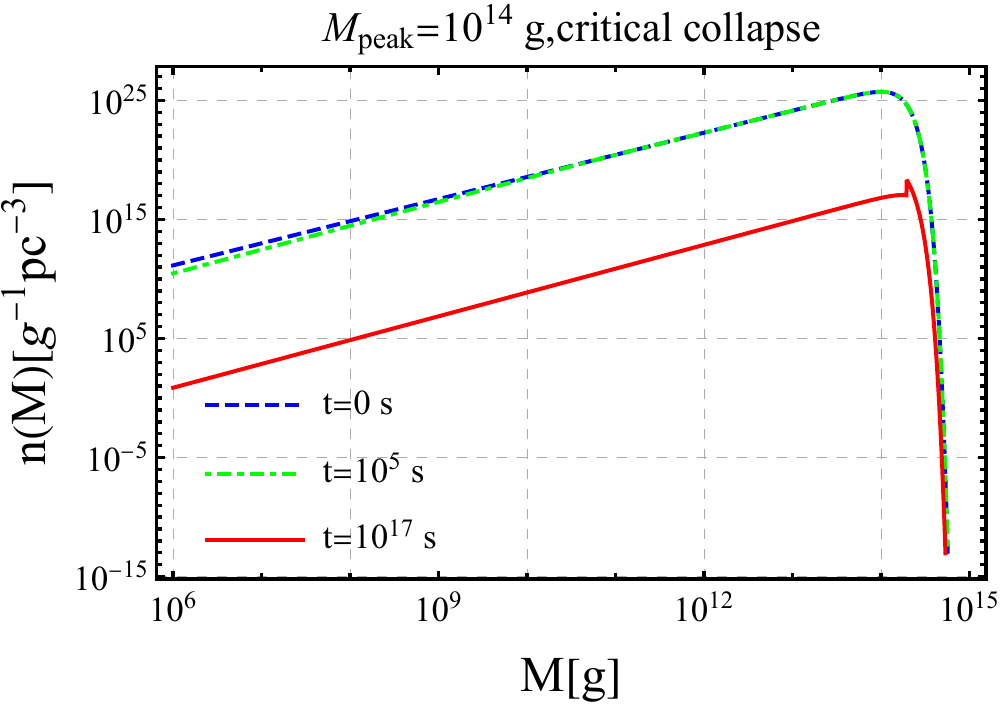}
	\caption{The mass functions at times $t = 0, 10^5, 10^{17}$ s for the initial critical collapse mass function, denoted by the blue dashed, green dashed and red solid curves, respectively. We choose the parameters for the initial mass function: $M_{pk} = 10^{14}$ g, $\beta = 10^{-23}$.}
	\label{fig:Crit}
\end{figure}

To check the above expressions, we take the case of critical collapse of PBH formation, i.e., the formation mass function is written as \cite{Carr:2020gox, Luo:2020dlg}
\be
n_c(M)
=
A(M_H) M^{1/\nu - 1} \exp\l[ - ( 1 - \nu ) \l( \frac{M}{ M_\text{peak} } \r)^{1/\nu} \r] ~.
\ee
For simplicity, we set $A=1$, $\nu = 0.35$ and $M_\text{peak} = 10^{14}$ g, respectively. Using the formulas \eqref{TimeEvolvingMassFunc} and \eqref{tranfer_func}, the time-evolving mass functions at various times are shown in Fig. \ref{fig:Crit}.
Note that the plots in Fig. \ref{fig:Crit} are more smooth than those of Ref. \cite{Carr:2016hva}, since we have used the more precise approximation to the mass-loss rate, see Eq. \eqref{phi_appro}.

\bibliographystyle{apsrev4-1}
\bibliography{pbh_timer}

\end{document}